\documentclass{article} 
\usepackage{iclr2025_conference,times}

\usepackage{amsmath,amsfonts,bm}

\def\eqref#1{equation~\ref{#1}}

\def\1{\bm{1}}

\DeclareMathAlphabet{\mathsfit}{\encodingdefault}{\sfdefault}{m}{sl}
\SetMathAlphabet{\mathsfit}{bold}{\encodingdefault}{\sfdefault}{bx}{n}

\usepackage{graphicx}
\usepackage{bbding}
\usepackage[normalem]{ulem}
\usepackage{multirow}
\usepackage{xcolor}
\usepackage{mdframed,lipsum}
\usepackage{hyperref}
\usepackage{url}
\usepackage{natbib}

\title{Edge-Enhanced Dilated Residual Attention Network for Multimodal Medical Image Fusion}

\author{Meng Zhou\thanks{Equal Contribution, corresponding to M.Zhou}, Yuxuan Zhang\footnotemark[1], Xiaolan Xu\\
Department of Computer Science\\
University of Toronto\\
Toronto, Canada\\
\texttt{\{simonzhou,yuxuan,landyxu\}@cs.toronto.edu} \\
\And
Jiayi Wang\\
Desautels Faculty of Management\\
McGill University\\
Montreal, Canada\\
\texttt{jiayi.wang10@mail.mcgill.ca} \\
\AND
Farzad Khalvati\\
Department of Computer Science, Department of Medical Imaging\\
University of Toronto\\
Toronto, Canada\\
\texttt{farzad.khalvati@utoronto.ca}
}

\iclrfinalcopy 
\begin{document}

\maketitle

\begin{abstract}
Multimodal medical image fusion is a crucial task that combines complementary information from different imaging modalities into a unified representation, thereby enhancing diagnostic accuracy and treatment planning. While deep learning methods, particularly Convolutional Neural Networks (CNNs) and Transformers, have significantly advanced fusion performance, some of the existing CNN-based methods fall short in capturing fine-grained multiscale and edge features, leading to suboptimal feature integration. Transformer-based models, on the other hand, are computationally intensive in both the training and fusion stages, making them impractical for real-time clinical use. Moreover, the clinical application of fused images remains unexplored. In this paper, we propose a novel CNN-based architecture that addresses these limitations by introducing a Dilated Residual Attention Network Module for effective multiscale feature extraction, coupled with a gradient operator to enhance edge detail learning. To ensure fast and efficient fusion, we present a parameter-free fusion strategy based on the weighted nuclear norm of softmax, which requires no additional computations during training or inference. Extensive experiments, including a downstream brain tumor classification task, demonstrate that our approach outperforms various baseline methods in terms of visual quality, texture preservation, and fusion speed, making it a possible practical solution for real-world clinical applications. The code will be released at \url{https://github.com/simonZhou86/en_dran}.
\end{abstract}

\section{Introduction}

Medical imaging plays an increasingly prominent role in clinical diagnosis, it aims to aggregate common and complementary information from different image modalities as well as integrate the information to generate more clearer images~\citep{10256252}. Medical image fusion can enhance crucial details of anatomy and tissue information from different image modalities and hence helps physicians and radiologists in accurate diagnosis of diseases, e.g., precise localization of tumor boundaries and tissues~\citep{chen2024mr} and effective radiotherapy treatments~\citep{safari2023medfusiongan,10256252}. 

Multimodal medical images provide organizations and structures from various aspects. For instance, Magnetic Resonance Imaging (MRI) offers high-resolution soft-tissue anatomical details, while computerized tomography (CT) scans reveal high-density information like bone structures and implants~\citep{10256252}. Positron Emission Tomography (PET) and Single-Photon Emission Computed Tomography (SPECT) highlight metabolic activity and blood flow in tissues. Due to hardware limitations, single-modal images cannot provide accurate and comprehensive information~\citep{10256252}. Physicians must analyze multiple images to make informed decisions, a process that is both time-consuming and laborious. Multimodal medical image fusion (MMIF) addresses this challenge by integrating prominent and complementary information into a single, more visually perceptive image, thereby supporting more accurate diagnoses~\citep{wang2022functional}.

Recent advances in deep learning have significantly improved multimodality fusion performance due to its powerful representation capabilities. Several works have utilized convolutional neural networks (CNNs) for various image fusion tasks. For example,~\cite{li2019dense} proposed DenseFuse, an infrared and visible image fusion framework using dense blocks. To address the limitation of single-scale feature extraction in~\cite{li2019dense},~\cite{song2019msdnet} introduced a multi-scale DenseNet (MSDNet), capturing features at different scales with various convolutional kernel sizes. Additionally,~\cite{zhang2020ifcnn} proposed a general CNN-based image fusion framework for multi-focus, infrared-visible, and multimodal medical image fusion. They introduced elementwise fusion rules to combine feature maps directly. There are also many efforts on solely multi-modal medical image fusion tasks.~\cite{fu2021multiscale} introduced the residual pyramid attention network capable of better deep feature extraction capabilities for MRI-CT, MRI-PET, and MRI-SPECT fusion tasks. They also proposed a Feature Energy Ratio Strategy to fuse two feature maps in the latent space. Similarly,~\cite{li2022multiscale} proposed a double residual attention network to capture detailed features while avoiding gradient vanishing or explosion. However, both~\cite{fu2021multiscale} and~\cite{li2022multiscale} suffer from losing the
structural information and edge details, which are crucial for medical images~\citep{meng2019mri}. Recently, Transformer-based methods~\citep{vaswani2017attention} have also garnered attention in computer vision tasks~\citep{dosovitskiy2020image}, with some applications in medical image fusion. For example,~\cite{ma2022swinfusion} proposed SwinFusion, combining a CNN feature extractor with a cross-domain transformer model to fuse local and global information, achieving superior fusion performance on MRI-CT and MRI-PET tasks. However, the high computational cost of computing global interactions hinders its clinical applications.~\cite{10256252} proposed a multiscale CNN model and applied residual Swin Transformer layers in the fusion strategy. Although their method further improved the fusion performance on all MRI-CT, MRI-PET, and MRI-SPECT tasks, the computational cost is still high and hinders the clinical application. Recently,~\cite{xie2024mactfusion} proposed a lightweight cross-modality transformer using window and grid attention, minimizing computational costs while maintaining superior fusion performance. More recently, the emergency of the improved selective structured state space models (Mamba)~\citep{gu2023mamba}, provides a novel solution to the problem above. Mamba has been demonstrated to outperform Transformer models in tasks requiring long-term dependency modeling, due to its selective global information modeling capabilities while maintaining linear complexity. This design not only reduces the computational costs but also enhances the inference speed. Notably, several studies have already explored the application of Mamba in various multimodal image fusion tasks~\citep{peng2024fusionmamba,li2024mambadfuse,xie2024fusionmamba}. Despite these advancements, many of the existing fusion methods lack validation on downstream tasks and remain unexplored. Therefore, a computationally friendly model that preserves both soft-tissue and detailed structural information (e.g., edges) is crucial for practical clinical applications. To this end, we propose a novel end-to-end feature fusion framework for multimodal anatomical and functional image fusion. We introduce the Dilated Residual Attention Network (DRAN) to extract multi-scale features and a family of \textit{parameter-free} fusion strategies to fuse feature maps in the latent space. Our approach combines input images to generate a fused image with more detailed textures and less information loss, surpassing several state-of-the-art fusion methods both qualitatively and quantitatively. We also take the clinical applicability into account and validate our approach on a downstream region-of-interest(ROI)-based brain tumor classification task between high-grade gliomas (HGG) and low-grade gliomas (LGG) following a similar procedure done in~\cite{zhou2024conditional}. To summarize, our contributions are as follows:
\begin{enumerate}
    \item We propose a novel Dilated Residual Attention Network (DRAN) for extracting multi-scale fine-grained features effectively. We also incorporate a learnable dense residual gradient operator (DRGO) to enhance edge feature representations.
    \item We introduce a family of \textit{parameter-free} fusion strategy, Softmax Feature Weighted Strategy, that outperforms other existing parameter-free fusion strategies and takes a step further to real-time fusion.
    \item To the best of our knowledge, we are the first to provide preliminary evaluations of the proposed fusion framework on a downstream brain LGG/HGG tumor type classification task.
    \item Experiments show our proposed method outperforms several baselines in both objective fusion metrics and subjective image quality, as well as the performance on the downstream classification task.
\end{enumerate}

\section{Materials and Method}

\begin{figure*}[h] 
	\begin{center}
		\includegraphics[width=\textwidth]{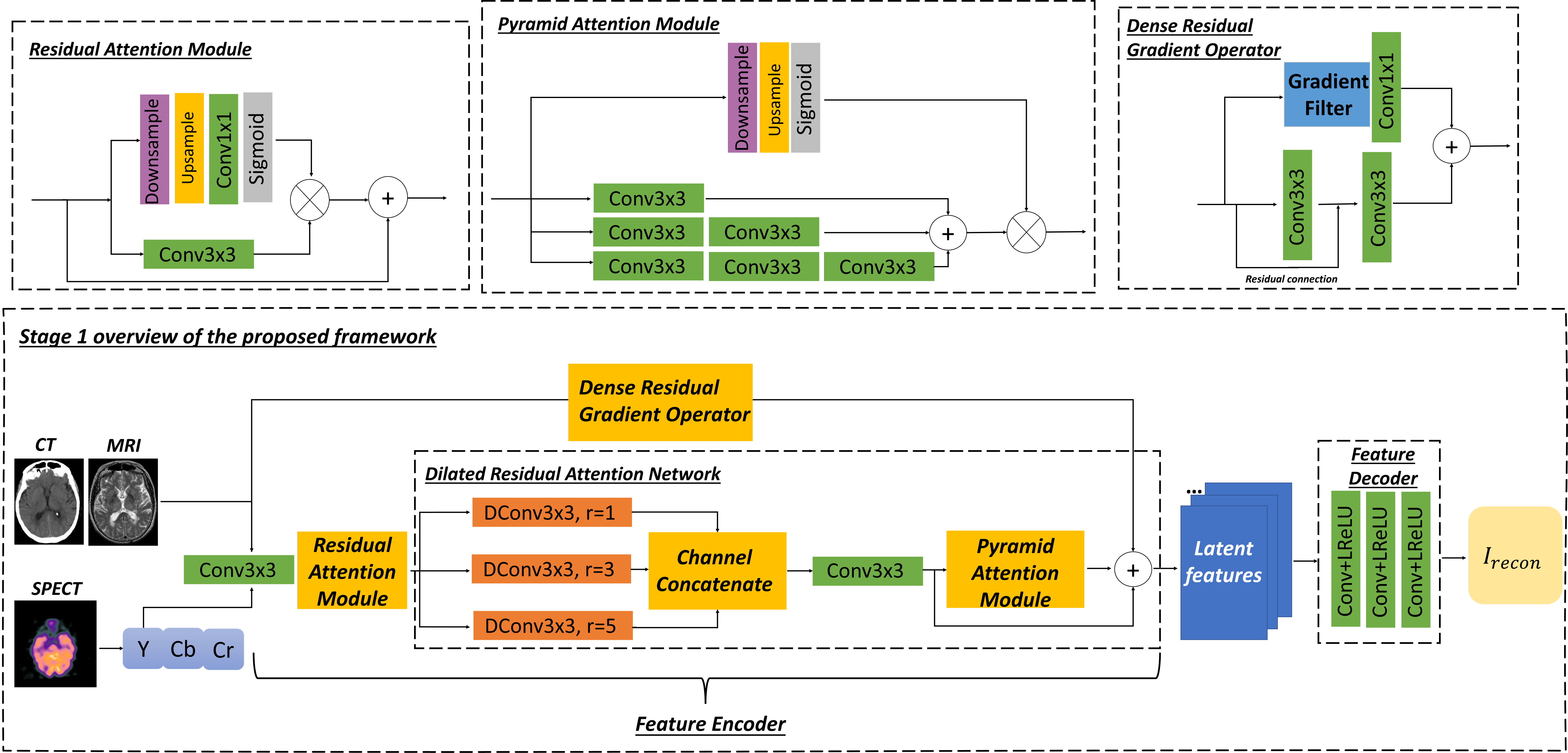}
	\end{center}
	\caption{An overview of Stage 1 model in the proposed framework. \texttt{DConv3x3, r=c} represents dilated convolution with kernel size 3$\times$3 and dilation rate equals to c. All \texttt{Conv+LReLU} layers in the decoder have 3$\times$3 kernel followed by Leaky-ReLU. Note that the YCbCr conversion only applies to SPECT images.}
	\label{overall}
\end{figure*}

\begin{figure*}[h] 
	\begin{center}
		\includegraphics[width=\textwidth]{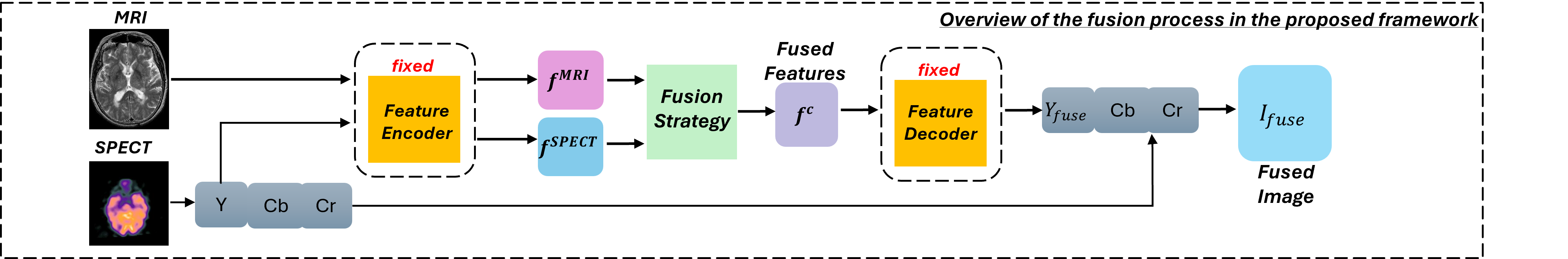}
	\end{center}
	\caption{A sample illustration of the fusion process, we take MRI-SPECT fusion as an example. The fusion process is the same for MRI-CT fusion except the RGB to YCbCr conversion is ignored.}
	\label{fusion_process}
\end{figure*}

The proposed fusion framework consists of three main components: a feature encoder, a fusion module, and a feature decoder. Specifically, we designed an \textbf{asymmetric autoencoder} comprising a deep feature encoder and a lightweight feature decoder. The autoencoder extracts multi-scale features from the input image into the latent space using the proposed DRAN and an edge enhancer, then reconstructs the image back to the original space, as illustrated in Figure~\ref{overall}. Since there is no golden ground truth for the fusion task, we disentangle the training process of the proposed fusion framework into two stages following recent works~\citep{10256252,fu2021multiscale,li2022multiscale}. In the first stage, we train the autoencoder \textit{solely} to extract multi-scale deep features and reconstruct images through a general reconstruction task. In the second stage depicted in Figure~\ref{fusion_process}, we introduce a family of \textit{parameter-free} fusion strategy utilizing softmax weights and the nuclear norm to fuse two feature maps, \textbf{$f^a$} and \textbf{$f^b$}, into a single feature map \textbf{$f^c$}. Using the trained decoder, this fused feature map \textbf{$f^c$} is then decoded back to the image space to obtain the final fused image. Below, we provide a detailed explanation of each component.

\subsection{Feature Extraction and Image Reconstruction}
\noindent \textbf{Feature Encoder. }The core module in the feature encoder is the \textbf{Dilated Residual Attention Network} (DRAN), which is inspired by two state-of-the-art mechanisms: residual attention~\citep{wang2017residual} and pyramid attention~\citep{li2018pyramid}. The residual attention mechanism includes a sequence of convolutional layers for feature processing and, in parallel, a downsample-upsample block with a sigmoid function for learning weights based on feature importance. The residual attention network enhances the feature expression capability through the attention mechanism and aids in faster model convergence by preventing gradient vanishing or explosion via the residual connection. A graphical illustration of our residual attention module is shown at the top-left of Figure~\ref{overall}. However, residual attention alone cannot effectively extract and learn multi-scale semantic features. To address this, we incorporate an additional pyramid attention network~\citep{li2018pyramid}. The convolution block in pyramid attention usually contains multiple convolutional layers that capture features at different scales and receptive fields. We adopt this approach by replacing convolutions with larger kernel filters with a sequence of smaller kernel filters~\citep{szegedy2016rethinking}. As shown at the top-middle of Figure~\ref{overall}, we use a single $3 \times 3$ convolution to represent a $3 \times 3$ receptive field, \textit{two} $3 \times 3$ convolutions to represent a $5 \times 5$ receptive field, and \textit{three} $3 \times 3$ convolutions to represent a $7 \times 7$ receptive field. Different from \cite{fu2021multiscale}, we leverage the ${\{1,3,5\}}$-dilated convolution~\citep{yu2015multi} on shallow features of the original input image to further enhance the learning ability of local multi-scale information and fine details without downsampling the feature map. The receptive field is expanded using three different dilated convolutions to improve the \textit{discriminative} multi-scale feature extraction ability of the model.  Once the multi-scale features are extracted, they are concatenated channel-wise. The residual-pyramid attention paradigm is then applied to further extract deep features. These deep features are the output of the feature extraction module and are used in both the fusion and reconstruction modules.

Edge information in MRI and CT is also crucial for accurate feature representation. To precisely capture these fine-grained edge features, we introduce a learnable dense residual gradient operator (DRGO) to enhance edge feature representation. The proposed module aggregates learnable convolutional features with gradient magnitude information on shallow features, and these are directly added element-wise to the features after the DRAN block, as shown at the bottom of Figure~\ref{overall}. Our DRGO module uses two convolutional layers with 3$\times$3 kernel sizes with residual connections to extract features. Additionally, a Sobel gradient operator~\citep{kanopoulos1988design}, followed by a 1$\times$1 convolutional layer, is used to learn the gradient information. This combination enhances the edge features by integrating both convolutional and gradient-based information. A graphical illustration of the DRGO module is shown at the top-right of Figure~\ref{overall}.

\noindent \textbf{Asymmetric Autoencoder. }The detailed architecture for the feature encoder is shown in Figure~\ref{overall}. After we obtained the latent features of input images from the feature encoder, we passed them into a lightweight decoder which only contains three convolutional layers and leaky ReLU activation function. The rationale behind this asymmetric design is that if the latent feature maps contain a wealth of useful information, they can be easily reconstructed back to the original image. Therefore, we employ a lightweight decoder to maximize the encoder's capability to extract rich semantic information in the latent space. Note that the latent features will be directly used in the subsequent fusion process, making the informative latent features crucial to the quality of the overall fused image.

\subsection{Feature Fusion} \label{fs}

\noindent \textbf{Softmax Weighted Fusion Strategy. }The fusion strategy is used to fuse the extracted features of input images into a single feature map. In this work, we introduce a novel parameter-free fusion strategy termed Softmax Feature Weighted Strategy. First, we obtained two output feature maps $f^{a}, f^{b}$ from the extraction module for input images $I_a, I_b$, respectively. These feature maps can be used to generate the corresponding weight maps that indicate the amount of contribution of each pixel to the final fused feature map~\citep{lahoud2019zero}. then, to get the weight map, we take the channel-wise softmax operation to the feature map which can be realized by Equation~\eqref{softmax_op}, where $x_i$ is the $i$th channel of the output feature map $x$.

\begin{equation} \label{softmax_op}
    S(x_i) = \frac{\exp(x_i)}{\sum_{i} \exp(x_i)}
\end{equation}

After we obtained the softmax output, we computed the matrix nuclear norm ($\|\cdot\|_{*}$), which is the summation of its singular values. Finally, we obtain the weights for the output feature map by taking the weighted average of the \textit{maximum value} of the nuclear norm. The formula is given in Equation \eqref{sfnn}.


\begin{equation} \label{sfnn}
    W_k = \frac{\phi(\|S(x_{i})^k\|_{*})}{\sum_{k=1}^{C} \phi(\|S(x_{i})^k\|_{*})}
\end{equation}

Where $C$ is the number of multimodality images ($C = 2$ in our work), $S(x_{i})^k, k \in [a,b]$ is the weight map after softmax operation for the feature map $f^k, k \in [a,b]$. As mentioned, we selected $\phi(\cdot)$ to be \texttt{max()} in this work, but it can also be \texttt{mean()}, \texttt{sum()} or \texttt{identity()}. The final fused feature map is then given by $f = \sum W_k * f_k, k \in [a,b]$.


\subsection{Loss Function}
Since only the autoencoder in Stage 1 involves training and optimization, we use the reconstruction objective which combines pixel loss, image gradient loss~\citep{ma2020structure, fabbri2018enhancing}, and perceptual loss~\citep{johnson2016perceptual} to optimize the network. The gradient loss~\citep{ma2020structure, fabbri2018enhancing} is added to model the fine details of textures in the reconstructed image, and the perceptual loss~\citep{johnson2016perceptual} is added to model the high-level semantic similarity between reconstructed and input images. In detail, our loss function is defined as follows:


\begin{equation} \label{l1gradperp}
    \begin{gathered}
        \mathcal{L}_{pixel} = \|x - \hat{x}\|_{2}^{2}, \ \mathcal{L}_{grad} = \|\nabla x - \nabla \hat{x}\|_{2}^{2}, \\
        \mathcal{L}_{perp} = \sum_{k=1}^{C} \|f_{i}^{k}(x) - f_{i}^{k}(\hat{x})\|_{2}^{2}
    \end{gathered}
\end{equation}

\begin{equation} \label{total_loss}
    \mathcal{L}(\theta) = \mathcal{L}_{pixel} + \lambda_1 * \mathcal{L}_{grad} + \lambda_2 * \mathcal{L}_{perp}
\end{equation}

Equation~\eqref{l1gradperp} shows the three losses utilized. We use a standard $\mathcal{L}_2$ distance for the pixel loss, where $M$ is the number of input images, $I_o$ is the output image. The image gradient loss is also realized by the $\mathcal{L}_2$ norm of the image gradient in both $x$ and $y$-direction. Finally, the perceptual loss~\citep{johnson2016perceptual}, where $f_{i}^k(x)$ is the the $k$-th channel in $i$-th layer (with size $W_i \times H_i$) from the pre-trained VGG16 network~\citep{simonyan2014very} with input image $x$, and $C$ is the number of channels. We prefer $i$ to be large, i.e., the deeper layer of the VGG network. The total loss function is given in Equation~\eqref{total_loss}, $\theta$ is the set of network weights to be optimized, $\lambda_1, \lambda_2$ are weight balancing factors of the gradient and perceptual loss, respectively. We empirically set $\lambda_1, \lambda_2$ equal to 0.5.

\subsection{Data and Preprocessing}
In this work, we use three datasets to validate the effectiveness of our proposed approach: MRI-CT (184 pairs) and MRI-SPECT (357 pairs) multi-modality fusion from The Harvard Whole Brain Atlas dataset\footnote{\url{https://www.med.harvard.edu/aanlib/}}, and additional FLAIR and T2 sequence data from the BraTS 2019 dataset~\citep{bakas2017advancing,bakas2018identifying,menze2014multimodal} (335 patients) to evaluate our method on the downstream brain tumor type classification task. For MRI-SPECT fusion, we converted SPECT images from the RGB color space to the YCbCr space following~\citep{fu2021multiscale,10256252,li2022multiscale}, using only the Y-channel images to train the model. For MRI-CT fusion, we used the original images as they are all single-channel grayscale images. All MRI-CT and MRI-SPECT image pairs were co-registered and preprocessed beforehand so that each pixel intensity is in the range of [0,255]. We further normalized the pixel intensity to [0,1].

For the T2 and FLAIR sequence data from BraTS, we first obtained the ROIs by multiplying the images with masks, then reshaped the data from 240$\times$240$\times$155 to 128$\times$128$\times$128, and normalized all pixel intensities to the range [0,1]. We converted the 3D data to 2D by slicing over the \textit{Axial plane} for each patient and only considered slices with at least 10\% non-zero pixels. 

\section{Experiments}

All programs were implemented in PyTorch and were trained on Google Colab and Compute Canada. For both MRI-CT and MRI-SPECT pairs, we trained the autoencoder for 100 epochs with an initial learning rate of 0.0001 and cosine decay to 3e-7, a mini-batch size of 4, and with the Adam optimizer~\citep{kingma2014adam}. We randomly held out 30 image pairs from the MRI-CT dataset and 50 pairs from the MRI-SPECT dataset as the standalone test set. To ensure the robustness of our model, we repeated our experiments three times and ensured that we had different test sets in each run. 

To assess the usability of our fusion framework, we conducted an ROI-based brain tumor classification task between LGG and HGG using the BraTS 2019 data as we discussed previously. First, we randomly held out 40 \textit{patients} (20 LGG and 20 HGG patients, 1152 slices in total) as a standalone test set. The rest of the data is used to train our model. We trained our fusion autoencoder for 25 epochs with a constant learning rate of 0.0001, a mini-batch size of 4, and with the Adam optimizer. We used our proposed SFNN-max fusion strategy to fuse T2 and FLAIR images. For the classification model, we used ResNet-50 for all experiments and trained with focal loss~\citep{lin2017focal} followed~\citep{zhou2024conditional}. We trained the model for 50 epochs with a constant learning rate of 0.001, a mini-batch size of 8, and an Adam optimizer. We ran the classification experiment for three trials with different train-validation splits to ensure the robustness and reliability of our findings.

\noindent \textbf{Baseline Model \& Comparison. }For comparison of image fusion results on MRI-CT and MRI-SPECT data, we considered five open-sourced state-of-the-art methods: IFCNN~\citep{zhang2020ifcnn}, MSRPAN~\citep{fu2021multiscale}, MSDRA~\citep{li2022multiscale}, SwinFusion~\citep{ma2022swinfusion} and MRSCFusion~\citep{10256252}, where the first three are CNN-based methods and the last two are Transformer-based methods. We rerun all models except MRSCFusion due to memory limitations, and we used the pre-trained weights of MRSCFusion from their official repository\footnote{\url{https://github.com/millieXie/MRSCFusion}}. For quantitative comparisons, we select five commonly used metrics in previous works~\citep{10256252,li2022multiscale,fu2021multiscale,chen2024mr,safari2023medfusiongan}: Peak signal-to-noise ratio (PSNR), Structural Similarity (SSIM)~\citep{wang2004image}, Feature Mutual Information~\citep{haghighat2011non}, Feature SSIM (FSIM)~\citep{zhang2011fsim}, and Information Entropy (EN). The downstream classification performance is evaluated using AUC, F1-Score, and Accuracy.

\section{Results and Discussions}
\subsection{Image Fusion Results}

\begin{figure*}[!ht] 
	\begin{center}
		\includegraphics[width=\textwidth]{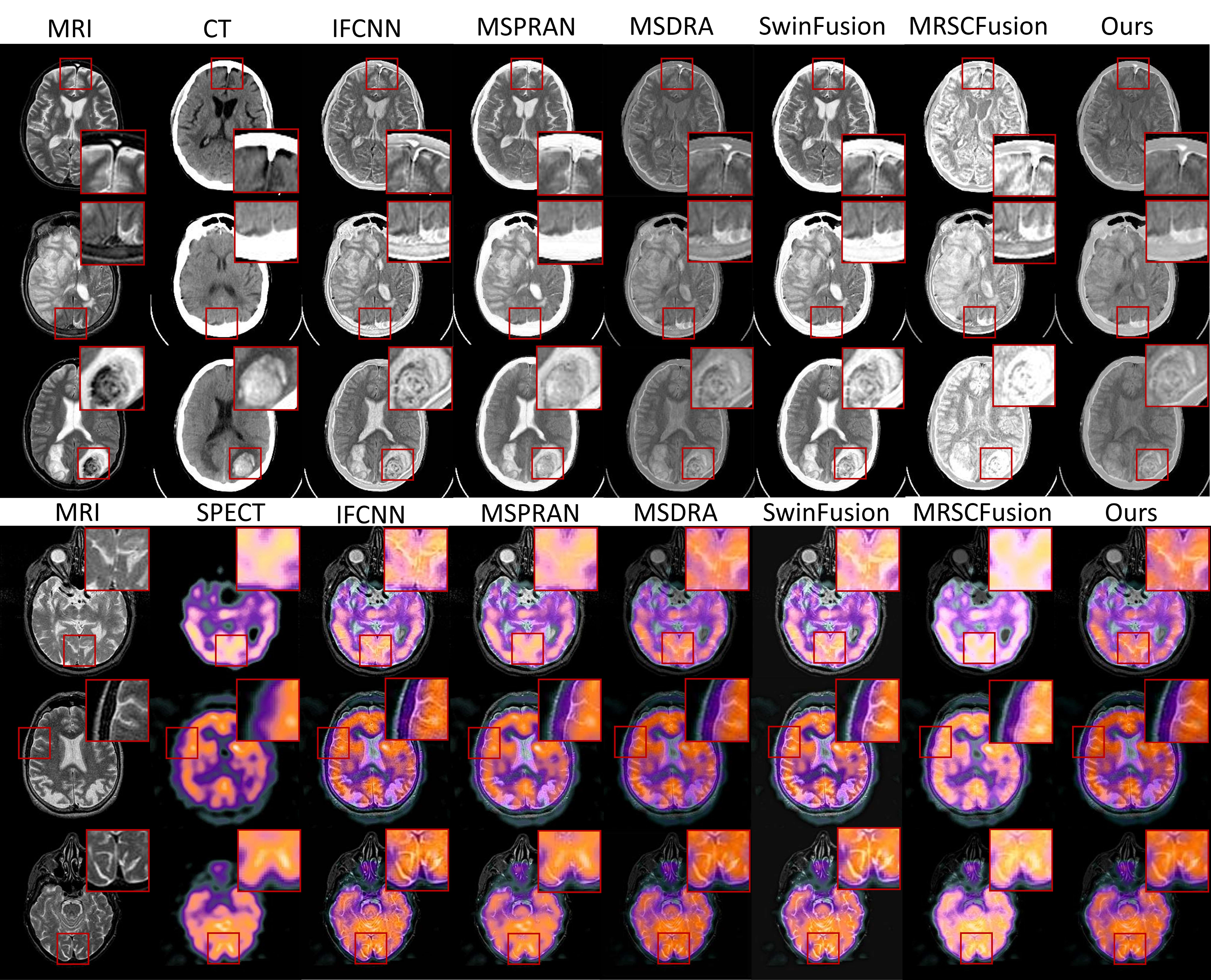}
	\end{center}
	\caption{Qualitative results for MRI-CT (top three rows) and MRI-SPECT (bottom three rows) fusion task. We randomly select three sample pairs from both test sets and show the fusion results across different methods. Zoom in for a better view.}
	\label{qual_res}
\end{figure*}

\noindent \textbf{Main Results. }In Figure~\ref{qual_res}, we compare the fusion results of different methods across three randomly selected MRI-CT and MRI-SPECT test pairs. For the MRI-CT fusion task, we focus on the area with more tissue information from MRI images and bone structures from CT images (highlighted and zoomed in red). We expect that the dense information (e.g., bone structures) in CT images and soft tissues in MRI images should be simultaneously retained in the fused images. We observed that the MRSCFusion had undesirable pixel intensities and some of the details from the MRI image were missing, such as the brain contour line. SwinFusion preserves tissue information from MRI images well but fails to maintain the boundary details for both CT and MRI images. IFCNN retains MRI tissue details well but slightly weakens dense structures from CT images. MSRPAN struggles to distinguish tissue boundaries between CT and MRI, losing fine-grained details (e.g., the second image in the MSRPAN column), resulting in overly sharp edges. MSDRA produces unclear boundaries and lacks significant intensity differences, making it difficult to differentiate between dense boundaries and soft tissues. In contrast, our proposed method provides a clear, bright brain contour from MRI and better edge contrast from CT. Our fused images offer superior contrast between edges and inner tissues, preserving more edge and fine-grained information from both modalities. Visually, our results appear more natural and the overall contrast is more promising.

For the MRI-SPECT fusion task, we focus on the area with more morphological (color) information from the SPECT and texture information from MRI images (highlighted and zoomed in red box) for a better comparison. MRSCFusion introduces noticeable intensity distortions and loses texture details from MRI images. SwinFusion achieves satisfactory results, but some color information blurs the MRI texture. IFCNN effectively represents the functional information from SPECT images but still misses some fine details from MRI images. MSRPAN and MSDRA preserve color information well but blur some MRI texture details. Our fused images retain the appropriate color information from SPECT while preserving more structure and tissue details from MRI.

The quantitative metrics, computed over three distinct test sets, are reported with mean values and standard deviations in Table~\ref{main_res}. Our EH-DRAN method achieves the best performance in terms of PSNR, FMI, FSIM, and Information Entropy for the MRI-CT fusion task. The high FMI, FSIM, and Information Entropy scores indicate that our fused images maintain superior structural similarity and contain richer information. Although our method shows a slightly lower SSIM score compared to SwinFusion, this may be attributed to SwinFusion's tendency to emphasize MRI features. In contrast, our approach balances the contributions from both MRI and CT images.

For the MRI-SPECT fusion task, our method consistently surpasses baseline methods in PSNR, SSIM, FMI, and FSIM. Despite slightly lower Entropy values than SwinFusion, all other metrics demonstrate that our approach effectively preserves more functional and morphological information from MRI and SPECT images. This outcome aligns with the qualitative fusion results discussed earlier.

\begin{table*}[h]
\caption{Comparison between different methods on two test datasets, \textbf{bold} and \underline{underline} numbers represent best and second-best values in each dataset, respectively.}
\centering
\resizebox{\columnwidth}{!}{
\begin{tabular}{ccccccc}
\hline
Dataset              & Method                                                   & PSNR                      & SSIM                     & FMI                      & FSIM                     & EN                   \\ \hline
MRI-CT               & IFCNN~\citep{zhang2020ifcnn}        & \underline{15.594$\pm$0.112}    & 0.700$\pm$0.015          & 0.870$\pm$0.012          & 0.801$\pm$0.001          & 8.968$\pm$0.227           \\
                     & MSRPAN~\citep{fu2021multiscale}     & 14.790$\pm$0.233          & 0.749$\pm$0.003          & 0.744$\pm$0.001          & 0.804$\pm$0.001          & 7.773$\pm$0.273           \\
                     & MSDRA~\citep{li2022multiscale}      & 15.308$\pm$0.437          & 0.742$\pm$0.037          & 0.872$\pm$0.002          & 0.788$\pm$0.005          & \underline{9.554$\pm$0.767}     \\
                     & SwinFusion~\citep{ma2022swinfusion} & 14.962$\pm$0.173          & \textbf{0.768$\pm$0.007} & \underline{0.882$\pm$0.002}    & \underline{0.810$\pm$0.001}    & 8.445$\pm$0.078           \\
                     & MRSCFusion~\citep{10256252}         & 14.476$\pm$0.205          & 0.713$\pm$0.012          & 0.877$\pm$0.006          & 0.791$\pm$0.010          & 7.544$\pm$0.232           \\
                     & EH-DRAN(Ours)                                            & \textbf{16.830$\pm$0.490} & \underline{0.753$\pm$0.007}    & \textbf{0.883$\pm$0.005} & \textbf{0.820$\pm$0.003} & \textbf{10.727$\pm$0.531} \\ \hline
MRI-SPECT            & IFCNN~\citep{zhang2020ifcnn}        & \underline{19.728$\pm$0.228}    & 0.721$\pm$0.025          & \underline{0.846$\pm$0.062}    & 0.783$\pm$0.027          & 10.167$\pm$0.429          \\
\multicolumn{1}{l}{} & MSRPAN~\citep{fu2021multiscale}     & 19.174$\pm$0.046          & 0.732$\pm$0.002          & 0.838$\pm$0.003          & 0.793$\pm$0.002          & 9.737$\pm$0.202           \\
\multicolumn{1}{l}{} & MSDRA~\citep{li2022multiscale}      & 19.662$\pm$0.165          & 0.725$\pm$0.003          & 0.839$\pm$0.003          & 0.794$\pm$0.003          & 10.784$\pm$0.447          \\
\multicolumn{1}{l}{} & SwinFusion~\citep{ma2022swinfusion} & 17.557$\pm$0.021          & 0.728$\pm$0.004          & 0.808$\pm$0.007          & \underline{0.819$\pm$0.011}    & \textbf{13.066$\pm$0.428} \\
\multicolumn{1}{l}{} & MRSCFusion~\citep{10256252}         & 18.412$\pm$0.211          & \underline{ 0.734$\pm$0.012}    & 0.827$\pm$0.009          & 0.814$\pm$0.006          & 9.87$\pm$0.600            \\
\multicolumn{1}{l}{} & EH-DRAN(Ours)                                            & \textbf{21.455$\pm$0.071} & \textbf{0.736$\pm$0.002} & \textbf{0.876$\pm$0.004} & \textbf{0.843$\pm$0.003} & \underline{11.970$\pm$0.538}  \\ \hline
\end{tabular}
}
\label{main_res}
\end{table*}

\noindent \textbf{Selection of Fusion Strategies. }As discussed in Section~\ref{fs}, our proposed fusion strategy offers several variants, including $\phi(\cdot) = \texttt{mean()}, \texttt{sum()}$, and $\texttt{max()}$. We conducted an extensive qualitative and quantitative analysis of these strategies for both the MRI-CT and MRI-SPECT fusion tasks, with the results summarized in Table~\ref{fusion_stra}. For comparison, we used FER~\citep{fu2021multiscale} and FL1N~\citep{li2022multiscale}, two parameter-free fusion strategies from prior research, as baselines.

Our results demonstrated that the proposed fusion strategy consistently outperformed the baseline methods across all quantitative metrics. Notably, the robustness of our approach was evident, as the performance varied only slightly across different $\phi(\cdot)$ functions. This robustness underscores the stability and reliability of our strategy. Detailed qualitative comparisons of different fusion strategies are provided in Appendix~\ref{app:fs}.

\begin{table*}[h]
\caption{Comparison between different fusion strategies on two test datasets.}
\centering
\resizebox{\columnwidth}{!}{
\begin{tabular}{ccccccc}
\hline
Dataset              & Fusion Strategy                                                   & PSNR                      & SSIM                     & FMI                      & FSIM                     & EN                   \\ \hline
MRI-CT               & FER~\citep{fu2021multiscale} & 14.718$\pm$0.549 & 0.743$\pm$0.005 & 0.874$\pm$0.002 & 0.798$\pm$0.008 & 8.675$\pm$0.182 \\
                     & FL1N~\citep{li2022multiscale} & 15.620$\pm$0.279 & 0.737$\pm$0.002 & 0.878$\pm$0.003 & 0.804$\pm$0.008 & 9.001$\pm$0.583 \\
                     & SFNN-mean & 15.631$\pm$0.290 & 0.736$\pm$0.003 & 0.877$\pm$0.002 & 0.820$\pm$0.003 & 9.013$\pm$0.351 \\
                     & SFNN-max & 16.830$\pm$0.490 & 0.753$\pm$0.007 & 0.883$\pm$0.005 & 0.820$\pm$0.003 & 10.727$\pm$0.531 \\
                     & SFNN-sum & 15.590$\pm$0.370 & 0.735$\pm$0.005 & 0.876$\pm$0.003 & 0.810$\pm$0.007 & 9.397$\pm$0.732 \\ \hline
MRI-SPECT            & FER~\citep{fu2021multiscale} & 19.635$\pm$0.039 & 0.832$\pm$0.004 & 0.832$\pm$0.005 & 0.796$\pm$0.003 & 9.751$\pm$0.112 \\
\multicolumn{1}{l}{} & FL1N~\citep{li2022multiscale} & 20.337$\pm$0.058 & 0.833$\pm$0.002 & 0.842$\pm$0.006 & 0.800$\pm$0.010 & 10.562$\pm$0.482 \\
\multicolumn{1}{l}{} & SFNN-mean & 20.337$\pm$0.043 & 0.734$\pm$0.003 & 0.841$\pm$0.003 & 0.836$\pm$0.006 & 10.115$\pm$0.574 \\
\multicolumn{1}{l}{} & SFNN-max & 21.455$\pm$0.071 & 0.736$\pm$0.002 & 0.876$\pm$0.004 & 0.843$\pm$0.003 & 11.970$\pm$0.538 \\
\multicolumn{1}{l}{} & SFNN-sum & 20.336$\pm$0.136 & 0.734$\pm$0.002 & 0.842$\pm$0.003 & 0.838$\pm$0.010 & 10.937$\pm$0.451  \\ \hline
\end{tabular}
}
\label{fusion_stra}
\end{table*}

\noindent \textbf{Fusion Time. }Next, we assess the model complexity by examining the total number of trainable parameters and the image fusion time for each image pair using the MRI-SPECT dataset, as detailed in Table~\ref{time_analy}. The MRI-SPECT dataset is selected due to its larger number of image pairs and its representation of a more complex task, which closely mirrors real-world scenarios. Compared to other baselines, our proposed method has a reasonable number of parameters and achieves image fusion in only ~1 second. The proposed parameter-free fusion strategy suggests great potential for \textbf{real-time} fusion in clinical settings. Furthermore, we expect the fusion time for MRI-CT pairs to be even shorter than that for MRI-SPECT, primarily due to the absence of color space conversion between RGB and YCbCr.

\begin{table}[h]
\renewcommand{\arraystretch}{1.3}
\caption{Comparison between different methods in average inference time on MRI-SPECT dataset using a T4 GPU.}
\centering
\resizebox{\columnwidth}{!}{
\begin{tabular}{ccccccc}
\hline & IFCNN & MSRPAN & MSDRA & SwinFusion & MRSCFusion & Ours \\
Params(M) & 0.08 & 0.10 & 0.20 & 0.97 & 23.00 & 0.50 \\
Time(s) & 0.89 & 0.79 & 0.81 & 1.31 & 2.85 & 1.26 \\
\hline
\end{tabular}
}
\label{time_analy}
\end{table}

\noindent \textbf{Ablation Study.} We performed our ablation studies in two folds. (1) to assess the effectiveness of the proposed Dense Residual Gradient Operator (DRGO) module to learn edge details, (2) to evaluate the impact of incorporating the gradient loss during model optimization. The ablation experiments were performed using both the MRI-CT and MRI-SPECT datasets. We hypothesize that (1) the edge enhancer aids the autoencoder in extracting meaningful edge features, thereby reinforcing edge details from both source images in the fused images, and (2) the gradient loss assists the model in learning and reproducing fine texture details in the reconstructed output. Table~\ref{abs_enhance} presents the results of our ablation study. We began with a baseline model that excluded both the DRGO module and the gradient loss ($\mathcal{L}_{grad}$). Adding $\mathcal{L}_{grad}$ to the loss function led to an improvement across all evaluation metrics compared to the baseline, confirming the benefit of gradient loss for optimizing texture detail learning. When we further integrated the DRGO module into the model, we observed additional improvements across all metrics. The notable increase in SSIM and FSIM scores demonstrates the DRGO module’s effectiveness in preserving edge information from both source images, thus validating its contribution to overall fusion quality.

\begin{table*}[h]
\caption{Ablations on the edge enhancer and loss function components of our proposed framework on both datasets. Base Model represents the model that is trained without using the DRGO module and the loss function only contains pixel and perceptual loss.}
\centering
\resizebox{\columnwidth}{!}{
\begin{tabular}{ccccccc}
\hline
Dataset              & Method                       & PSNR                      & SSIM                     & FMI                      & FSIM                     & Entropy                   \\ \hline
MRI-CT               & Base Model                         & 15.623$\pm$0.032          & 0.745$\pm$0.013          & 0.878$\pm$0.003          & 0.802$\pm$0.003          & 9.122$\pm$0.706           \\
                     &  Base Model+$\mathcal{L}_{grad}$               & 16.355$\pm$0.038          & 0.749$\pm$0.010          & 0.881$\pm$0.002          & 0.818$\pm$0.002          & 9.771$\pm$0.528           \\
                     & Base Model+$\mathcal{L}_{grad}$+DRGO & \textbf{16.830$\pm$0.490} & \textbf{0.753$\pm$0.007} & \textbf{0.883$\pm$0.005} & \textbf{0.820$\pm$0.003} & \textbf{10.727$\pm$0.531} \\ \hline
MRI-SPECT            & Base Model                         & 20.698$\pm$0.002          & \textbf{0.743$\pm$0.008} & 0.833$\pm$0.006          & 0.836$\pm$0.005          & 10.010$\pm$0.563          \\
\multicolumn{1}{l}{} & Base Model+$\mathcal{L}_{grad}$               & 20.738$\pm$0.026          & 0.740$\pm$0.011          & 0.837$\pm$0.002          & 0.838$\pm$0.004          & 10.454$\pm$0.426          \\
\multicolumn{1}{l}{} & Base Model+$\mathcal{L}_{grad}$+DRGO & \textbf{21.455$\pm$0.071} & 0.736$\pm$0.002          & \textbf{0.876$\pm$0.004} & \textbf{0.843$\pm$0.003} & \textbf{11.970$\pm$0.538}
\\ \hline
\end{tabular}
}
\label{abs_enhance}
\end{table*}

\subsection{Classification Results}

To validate the efficacy of the proposed fusion framework in clinical multi-modality brain tumor images, we trained a classification model to validate the efficacy of the proposed fusion framework in distinguishing between HGG and LGG brain tumor types. Following~\citep{xie2024mactfusion}, we used T2 and FLAIR images for the classification task. For comparison, we evaluated three approaches: single-modality images (using either T2 or FLAIR), dual-modality images (using a channel-wise concatenation of T2 and FLAIR), and T2-FLAIR fused images. The classification results, presented in Table~\ref{classi_res}, demonstrate that utilizing the fused image generated by our proposed fusion framework significantly improves performance in terms of AUC and F1-Score compared to the single-modality and dual-modality baselines. This improvement highlights our method's ability to enhance details in the brain tumor ROIs while preserving overall image contrast, corroborating the findings from the previous section. These results further suggest the potential of leveraging fused images for improved accuracy in real clinical diagnosis.

\begin{table}[h]
\renewcommand{\arraystretch}{1.3}
\caption{Comparison of brain tumor type LGG/HGG classification performance between single-modality, dual-modality, and T2-FLAIR fused images. Values are reported as mean$\pm$standard deviation.}
\centering
\begin{tabular}{ccccc}
\hline & AUC & F1-Score & Accuracy \\
T2 (1-channel) & 0.722$\pm$0.021 & 0.703$\pm$0.018 & 0.604$\pm$0.037 \\
FLAIR (1-channel) & 0.727$\pm$0.024 & 0.701$\pm$0.008 & 0.611$\pm$0.017 \\
T2+FLAIR (2-channel) & 0.723$\pm$0.028 & 0.717$\pm$0.012  & 0.640$\pm$0.015 \\
Fused (1-channel) & \textbf{0.769$\pm$0.003} & \textbf{0.723$\pm$0.006}  & \textbf{0.640$\pm$0.011} \\
\hline
\end{tabular}
\label{classi_res}
\end{table}

\section{Conclusions}
\noindent In this work, we proposed a novel asymmetric autoencoder architecture incorporating a Dilated Residual Attention Network (DRAN) for effective multi-scale feature extraction. Additionally, we integrated a Dense Residual Gradient Operator (DRGO) as an edge enhancer to capture fine-grained edge details. We introduced a family of \textit{parameter-free} softmax-weighted fusion strategies for multimodal image fusion, designed to operate without requiring parameter computation during both training and inference phases, pushing us a step further to achieve real-time image fusion. Our extensive evaluations demonstrate that the proposed method outperforms several baseline approaches in subjective visual quality and objective fusion metrics. The improved performance in the downstream brain tumor classification task further highlights the effectiveness of our fusion framework. We envision our approach being applied to subsequent disease localization tasks, radiotherapy treatment planning, and surgical navigation in real-world clinical settings.

In the future, we plan to extend our method from 2D to 3D medical image fusion, as 3D data is more common in practical medical imaging. Furthermore, we aim to explore the integration of Transformers or Selective State Spaces Models~\citep{gu2023mamba} to enhance feature extraction and image fusion capabilities.

\section*{Acknowledgment}

No funding was received for conducting this study. This research is enabled in part by support provided by Compute Canada (\url{https://alliancecan.ca/en}).


\bibliography{iclr2025_conference}
\bibliographystyle{iclr2025_conference}

\appendix
\section{More on Fusion Strategies} \label{app:fs}

\begin{figure*}[!ht] 
	\begin{center}
		\includegraphics[width=\textwidth]{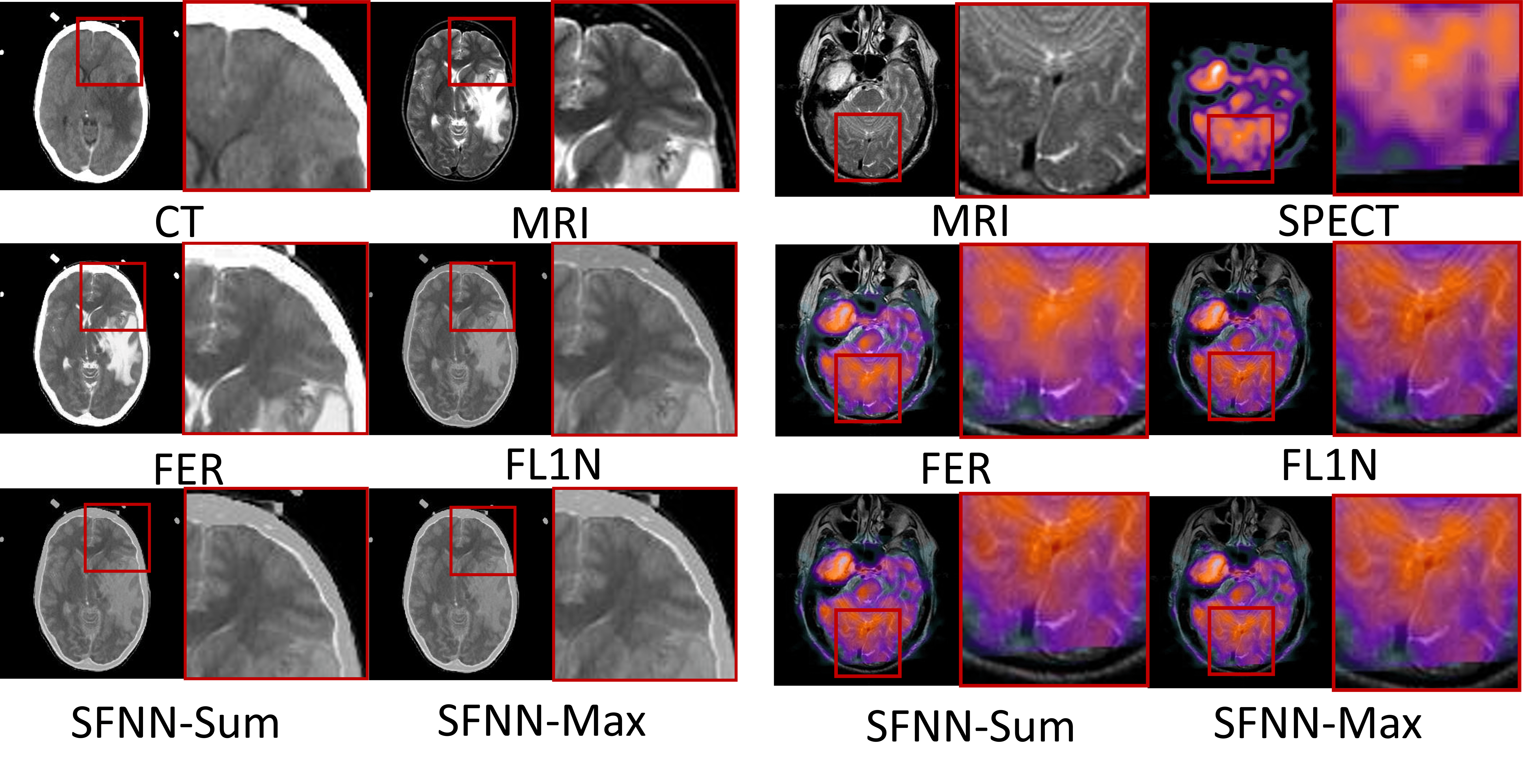}
	\end{center}
	\caption{Qualitative comparison between different fusion strategies on MRI-CT and SPECT dataset. \textit{Left panel:} visualization on MRI-CT dataset, \textit{right panel:} visualization on the MRI-SPECT dataset. For both datasets, we randomly sample a pair from the test set and zoom in on a selected region for a better view.}
	\label{qual_fs}
\end{figure*}

\noindent \textbf{Qualitative of different fusion strategies: }We selected two variants of our proposed SFNN methods, \texttt{sum()} and \texttt{max()}, due to their superior visual quality and performance, as illustrated in Figure~\ref{qual_fs}. For the MRI-CT fusion task, our method demonstrated clear advantages over the FER strategy~\citep{fu2021multiscale}, achieving enhanced fidelity with sharper delineation of inner tissue structures from MRI images and well-preserved boundaries from CT images. Furthermore, compared to the FL1N method~\citep{li2022multiscale}, our approach produced brighter and more defined edges, thereby improving the contrast between edges and tissues. In the case of MRI-SPECT fusion, our method retained more detailed features from the MRI image compared to the FER method. While the visual distinctions between our method and the FL1N strategy were subtle, the quantitative analysis further demonstrates the superiority of our approach.

\end{document}